From cwi.nl!paulv Fri May 22 13:45:25 1998
Return-Path: <paulv@cwi.nl>
Received: from hera.cwi.nl([192.16.191.1]) (45763 bytes) by maccs.dcss.mcmaster.ca
	via sendmail with P:esmtp/D:user/T:local
	(sender: <paulv@cwi.nl> owner: <real-jiang>) 
	id <m0ycvsd-0005tyC@maccs.dcss.mcmaster.ca>
	for <jiang@maccs.dcss.mcmaster.ca>; Fri, 22 May 1998 13:45:19 -0400 (EDT)
	(Smail-3.2.0.101 1997-Dec-17 #1 built 1997-Dec-22)
Received: from gnoe.cwi.nl (gnoe.cwi.nl [192.16.201.252]) by hera.cwi.nl with ESMTP
	id TAA06096 for ; Fri, 22 May 1998 19:45:16 +0200 (MET DST)
Received: by gnoe.cwi.nl 
	id TAA19448; Fri, 22 May 1998 19:45:14 +0200 (MET DST)
Date: Fri, 22 May 1998 19:45:14 +0200 (MET DST)
From: Paul.Vitanyi@cwi.nl
Message-Id: <UTC199805221745.TAA19448.paulv@gnoe.cwi.nl>
To: Harry.Buhrman@cwi.nl, jiang@maccs.dcss.mcmaster.ca, mli@wh.uwaterloo.ca
Subject: Part II
Status: R
Content-Length: 45163

I enclose the draft of the paper as it is now. We need to scrutinize
thedobtful parts:

1) improvement of lower bound Com Compl to n
2) the randomize Com Compl result.

I will not touch it over the weekend, so it is all yours.

Harry is going to STOC so he is out of it for a couple of days.

Cheers,  Paul

\documentstyle[11pt]{article}
\input{psfig}

\newtheorem{theorem}{\sc Theorem}
\newtheorem{lemma}{\sc Lemma}
\newtheorem{coro}{\sc Corollary}
\newtheorem{nota}{\sc Notation}
\newtheorem{defin}{\sc Definition}
\newtheorem{rem}{\sc Remark}
\newtheorem{cla}{\sc Claim}
\newtheorem{ex}{\sc Example}

\newenvironment{proof}{\par \sc Proof.\rm}{\hspace*{\fill}$\Box$\vspace{1ex}}

\newenvironment{definition}{\begin{defin}}{\end{defin}}
\newenvironment{remark}{\begin{rem}}{\end{rem}}

\title{New Applications of the Incompressibility Method: Part II}
\author{Harry Buhrman\thanks{
Partially supported by the European Union
through NeuroCOLT ESPRIT Working Group Nr. 8556,
and by  NWO through NFI Project ALADDIN
number NF 62-376.
Address: CWI,
Kruislaan 413, 1098 SJ Amsterdam, The Netherlands.
Email: buhrman@cwi.nl}\\
CWI
\and
Tao Jiang\thanks{Supported in part by the NSERC
Research Grant OGP0046613 and a CGAT grant.
Address: Department of Computer Science,
McMaster University, Hamilton, Ont L8S 4K1, Canada.
Email: jiang@maccs.mcmaster.ca}\\
McMaster University
\and
Ming Li\thanks{
Supported in part by
the NSERC Research Grant OGP0046506, CITO, a CGAT grant, and the
Steacie Fellowship. Address:
Department of Computer Science, University of Waterloo,
Waterloo, Ont. N2L 3G1, Canada. E-mail: mli@math.uwaterloo.ca}\\
University of Waterloo
\and
Paul Vit\'{a}nyi\thanks{
Partially supported by the European Union
through NeuroCOLT ESPRIT Working Group Nr. 8556,
and by  NWO through NFI Project ALADDIN
number NF 62-376.
Address: CWI,
Kruislaan 413, 1098 SJ Amsterdam, The Netherlands.
Email: paulv@cwi.nl}\\
CWI and University of Amsterdam}
\date{}
 
\begin{document}
\maketitle
 
\begin{abstract}
The incompressibility method is an elementary yet powerful 
proof technique. It has been used successfully in many
areas~\cite{LiVi93}. To further
demonstrate its power and elegance
we exhibit new simple 
proofs using the incompressibility method.
\end{abstract}
 
\section{Introduction}
The incompressibility of individual random objects 
yields a simple but powerful proof technique:
{\em the incompressibility method}.
This method is a general purpose
tool that can be used to prove lower bounds on computational
problems, to obtain combinatorial properties of concrete objects, and
to analyze the average complexity of an algorithm. Since the early
1980's, the incompressibility method has been successfully used 
to solve many well-known questions that had been open for a
long time and to supply new
simplified proofs for known results. A survey is \cite{LiVi93}.

The purpose of this paper is pragmatic, in
the same style as \cite{LiVi94,JLV98}. We want to
further demonstrate how easy the incompressibility method can be used, via a 
new collection of simple examples. The proofs we have chosen to be
included in \cite{JLV98} and 
here are not difficult ones. They are from diverse
topics and most of these topics are well-known. 
Some of our results are new (but this is not important), and
some are known before. In all cases, the new 
proofs are much simpler than the old ones (if they exist).

\section{Kolmogorov Complexity and the Incompressibility Method}
We use the following notation.
Let $x$ be
a finite binary string. Then $l(x)$ denotes the {\em length}
(number of bits) of $x$. In particular, $l(\epsilon)=0$
where $\epsilon$ denotes the {\em empty word}.

We can map $\{0,1\}^*$ one-to-one onto the
natural numbers by associating each string with its index
in the length-increasing lexicographical ordering
\begin{equation}
( \epsilon , 0),  (0,1),  (1,2), (00,3), (01,4), (10,5), (11,6),
\ldots .
\label{(2.1)}
\end{equation}
This way we have a binary representation for the
natural numbers that is different from the standard
binary representation.
It is convenient not to distinguish between the
first and second element of the same pair,
and call them ``string'' or ``number''
arbitrarily. As an example, we have $l(7)=00$.
Let $x,y, \in {\cal N}$, where
${\cal N}$ denotes the natural
numbers. 
Let $T_0 ,T_1 , \ldots$ be a standard enumeration
of all Turing machines.
Let $\langle \cdot ,\cdot \rangle$ be a standard one-one mapping
from ${\cal N} \times {\cal N}$
to ${\cal N}$, for technical reasons chosen such that
$l(\langle x ,y \rangle) = l(y)+O(l(x))$.

Informally, the Kolmogorov complexity, \cite{Ko65},
of $x$ is the length of the
{\em shortest} effective description of $x$.
That is, the {\em Kolmogorov complexity} $C(x)$ of
a finite string $x$ is simply the length
of the shortest program, say in
FORTRAN (or in Turing machine codes)
encoded in binary, which prints $x$ without any input.
A similar definition holds conditionally, in the sense that
$C(x|y)$ is the length of the shortest binary program
which computes $x$ on input $y$. 
Kolmogorov complexity is absolute in the sense
of being independent of the programming language,
up to a fixed additional constant term which depends on the programming
language but not on $x$. We now fix one canonical programming
language once and for all as reference and thereby $C()$.
For the theory and applications, as well as history, see \cite{LiVi93}.
A formal definition is as follows:

\begin{definition}
\rm
Let $U$ be an appropriate universal Turing machine
such that 
\[U(\langle \langle i,p \rangle ,y \rangle ) =
T_i (\langle p,y\rangle) \]
 for all $i$ and $\langle p,y\rangle$.
The {\em conditional Kolmogorov complexity} of $x$ given $y$
is
\[C(x|y) = \min_{p \in \{0,1\}^*} \{l(p): U (\langle p,y\rangle)=x \}. \]
The unconditional Kolmogorov complexity of $x$ is defined
as $C(x) := C(x| \epsilon )$.
\end{definition}
It is easy to see that there are strings that can be described
by programs much shorter than themselves. For instance, the
function defined by $f(1) = 2$ and $f(i) = 2^{f(i-1)}$
for $i>1$ grows very fast, $f(k)$ is a ``stack'' of $k$ twos.
Yet for each $k$ it is clear that $f(k)$
has complexity at most $C(k) + O(1)$.

By a simple counting argument one can show
that whereas some strings can be enormously compressed,
the majority of strings can hardly be compressed
at all.
For each $n$ there are $2^n$ binary
strings of length $n$, but only
$\sum_{i=0}^{n-1} 2^i = 2^n -1$ possible shorter descriptions.
Therefore, there is at least one binary string
$x$ of length $n$ such that $C(x)   \geq   n$.
We call such strings $incompressible$. It also
follows that for any length $n$ and any binary string $y$,
there is a binary string $x$ of length $n$ such that
$C(x| y)   \geq   n$.
\begin{definition}
\rm
For each constant $c$ we say a string $x$ is
\it c-incompressible
\rm if $C(x)   \geq   l(x) -c$.
\label{string!incompressibility of}
\end{definition}
 
Strings that are incompressible (say, $c$-incompressible
with small $c$) are patternless,
since a pattern could be used to reduce
the description length. Intuitively, we
think of such patternless sequences as being random, and we
use ``random sequence'' synonymously with ``incompressible sequence.''
It is possible to give a rigorous formalization of the intuitive notion
of a random sequence as a sequence that passes all
effective tests for randomness, see for example \cite{LiVi93}.
 
How many strings of length $n$ are $c$-incompressible?
By the same counting argument we find that the number
of strings of length $n$ that are $c$-incompressible
is at least $2^n - 2^{n-c} +1$. Hence
there is at least one 0-incompressible string of length $n$,
at least one-half of all strings of length $n$ are 1-incompressible,
at least three-fourths  of all strings
of length $n$ are 2-incompressible, \ldots , and
at least the $(1- 1/2^c )$th part
of all $2^n$ strings of length $n$ are $c$-incompressible. This means
that for each constant $c   \geq   1$ the majority of all
strings of length $n$ (with $n   >   c$) is $c$-incompressible.
We generalize this to the following simple but extremely
useful %
{\it Incompressibility Lemma}.
\begin{lemma}
\label{C2}
Let $c$ be a positive integer.
For each fixed $y$, every
set $A$ of cardinality $m$ has at least $m(1 - 2^{-c} ) + 1$
elements $x$ with $C(x| y)   \geq   \lfloor \log m \rfloor  - c$.
\end{lemma}
\begin{proof}
By simple counting.
\end{proof}

As an example, set $A =  \{ x: l(x) = n  \}  $. Then the cardinality
of $A$ is $m = 2^n$.
Since it is easy to assert that $C(x)  \leq n + c$ for some
fixed $c$ and all $x$ in $A$, Lemma~\ref{C2} demonstrates
that this trivial estimate is quite sharp. The deeper
reason is that since there are few short programs, there can
be only few objects of low complexity.

\begin{definition}
\rm
  A {\em prefix set}, or prefix-free code, or prefix code, is a set of
strings such that no member is a prefix of any other member.
  A prefix set which is the domain of a partial recursive function
(set of halting programs for a Turing machine) is a special type of
prefix code called a {\em self-delimiting} code because there is an
effective procedure which reading left-to-right 
determines where a code word ends without
reading past the last symbol.
  A one-to-one function with a range that is a self-delimiting code
will also be called a self-delimiting code.
\end{definition}

A simple self-delimiting code we use throughout is obtained by reserving one
symbol, say 0, as a stop sign and encoding a
natural number $x$ as $1^x 0$.
We can prefix an object with its length and iterate
this idea to obtain ever shorter codes:
\begin{equation}
\label{ladder}
E_i (x)  = \left\{ \begin{array}{ll}
1^x 0 & \mbox{for $i=0$}, \\
E_{i-1} (l(x)) x & \mbox{for $i>0$}.
\end{array} \right.
\end{equation}
Thus, $E_1 (x) = 1^{l(x)} 0 x$ and has
length $l(E_1 (x)) = 2l(x) + 1$; 
$E_2 (x)  =  \lg_1 (l(x)) x$ and has length
$ l(E_2 (x) )  =  l(x) + 2l(l(x)) + 1$.
   We have for example
 \[
  l(E_3 (x))\leq l(x)+\log l(x) + 2\log\log l(x) + 1.
 \]

  Define the pairing function
 \begin{equation}\label{e.ngl}
  \langle x,y \rangle =E_2(x)y
 \end{equation}
  with inverses $\langle \cdot \rangle _1,\langle\cdot\rangle_2$.
This can be iterated to
$\langle  \langle \cdot , \cdot \rangle , \cdot \rangle$.

In a typical proof using the incompressibility method,
one first chooses an individually random object from the
class under discussion.
This object is effectively incompressible.
The argument invariably says that if a desired property
does not hold, then the object
can be compressed. This yields the required contradiction.
Then, since most objects are random, the desired property 
usually holds on average.

\section{Number of Strings of Maximum Complexity}
A simple counting argument shows that for every $n$ there
is at least one string $x$ of length $n$ such that $C(x|n) \geq n$
and a string $y$ of length $n$ such that $C((y) \geq n$. In fact,
we can do much better.
With respect to the prefix version $K(\cdot)$
of Kolmogorov complexity 
reference \cite{Ch93} gives an elegant proof
that the number strings of length
$n$ that have maximal prefix complexity (also known as
self-delimiting complexity
or program-size complexity) is $\Omega (2^n)$.
\footnote{Prefix complexity makes its brief and only appearance in this
paper here; for more details check out
\cite{LiVi93}. We remark that the prefix complexity
$K(x)$ is typically larger than $C(x)$ and in fact for every $n$
there are $x$ such that $K(x)=n+K(n)+O(1)$. This is larger than
$C(x)$ which is upper bounded by $n+O(1)$. With respect to the
the related question for
$C(x)$ complexity \cite{Ch93} states: ``An earlier, unpublished
version of this result was obtained more than twenty years ago
in connection with 
[$C(\cdot )$]. $\ldots$ the proof shows
that a number is large because it is random.'' }

The purpose 
of this section is analyse this matter in detail
with respect to $C(\cdot)$ complexity. The situation is different
from prefix-complexity because here we have a simple constructive upper bound.
We also want to determine how large the $C(\cdot)$-complexity in fact
can get (and how many such strings there are). That these matters are not mere 
curiosities but can be used to obtain meaningful results are shown 
in \cite{JLV98}.
\footnote{
The history of interest and 
reinvention of these curious but useful
facts makes it useful to archive them.
Theorems~\ref{theo.max1}, \ref{theo.nurand2} were independently 
proved by two of us [HB,PV]
in June 1995, and Theorem~\ref{theo.max1} was also independently
found by
both M. Kummer
and L. Fortnow. This was not published but appears as Exercise 2.2.6
in \cite{LiVi93}. Of course, the cited reference \cite{Ch93}
giving the result for the prefix-complexity $K(\cdot)$ 
preceeds all of this. 
}
 
\begin{theorem}\label{theo.max1}
There is a constant $d>0$
such that for every $n$ there are at least $\lfloor 2^n/d \rfloor $
strings $x$ of
length $n$ with $C(x|n) \geq n$ (respectively, $C(x) \geq  n$).
\end{theorem}

\begin{proof}
It is well-known that there is a constant $c\geq0$
such that for every $n$ there is a string $x$
of length $n$ such that $C(x|n) \leq n+c$. Hence
for every $n$ and
every $x$ of length
$l(x) \leq n-c-1$ we have $C(x|n) < n$.
Consequently, there are at most $2^n - 2^{n-c}$
programs of length  $<n$ available as shortest programs for the strings
of length $n$ (there are $2^{n}-1$ potential programs and
$2^{n-c}-1$ thereoff are already taken). 
Hence there are at least $2^{n-c}$ strings $x$
of length $n$ with $C(x|n) \geq n$.
\end{proof}
 
\begin{theorem}\label{theo.nurand2}
There are constants $c,d>0$
such that for every large enough $n$
there are at least $\lfloor 2^n/d\rfloor$ strings $x$ of
length $n-c \leq l(x) \leq n$ with $C(x|n) > n$ (respectively, $C(x) >  n$).
\end{theorem}

\begin{proof}
For every $n$
there are equally many strings of length $\leq n$ to
be described and potential programs of length $\leq n$ to describe
them. Since
some programs do not halt 
for every large enough $n$ there exists a string $x$ of length at most $n$
such that $n < C(x|n) \leq l(x)+c$ and a string $y$ of length
at most $n$ such that $n < C(y) \leq l(y)+c$.

Let there be $m \geq 1$ such strings.
Given $m$ and $n$ we can enumerate
all $2^{n+1} - m -1$ strings $x$ of length $\leq n$ and
complexity $C(x|n) \leq  n$
by dovetailing the running
of all programs of length $\leq n$.
The lexicographic first string of length $ \leq n$
not in the list, say $x$, is described by a program $p$
 giving $m$ in $\log m$ bits
plus an $O(1)$-bit program to do the decoding of $x$. Therefore,
$\log m + O(1) \geq C(x|n) > n$ which proves the theorem
for the conditional case. The unconditional
result follows similarly by padding the description 
of $x$ up to length $n+c'$ for a constant $c'$ and adding
the description of $c'$ to program $p$ describing $x$. This way we can
first retrieve $c'$ from $p$
and then retrieve $n$ from the length of $p$.
\end{proof}
 
\begin{remark}
\rm
This shows that there are lots of strings $x$ that have complexity
larger than their lengths. How much larger can this get? While
the theorems above are invariant with respect to the choice
of the particular reference universal Turing machine
in the definition of the Kolmogorov complexity, the excess of
maximal complexity over the length depends on this choice. 

For example, we can easily choose a reference universal Turing machine
that has no halting programs of odd length, or such that
it has no halting programs of length $i \bmod 100$ for $i=0, \ldots, 98$.
In such a case there are many $x$'s that have complexity at least
$l(x)+100$. In the opposite extreme, given an appropriate universal Turing
machine $U$ we can transform it into a universal Turing
machine $U'$ such that $U'(1p)= p$ and $U'(0p)=U(p)$ for
all $p$. Taking $U'$ as reference universal Turing machine
we clearly have $C(x) \leq l(x)+1$ for all $x$.
Consequently, for every $n$ the shortest programs of strings of
length $<n$ have length at most $n$. This means that there are 
at most  $2^{n+1}-2^n = 2^n$ strings available of length
at most $n$ to serve as shortest programs for strings of length $n$.

By definition of $U'$ the Theorem~\ref{theo.nurand2} means
that at least $\Omega (2^n)$ strings of length $n$ are used as shortest
programs for strings of length $n-1$, while by definition
of $U'$ at least $2^n/2$ strings
of length $n$ are used as (not necessarily shortest)
programs for strings of length $n-1$.
Consequently, at most $2^n/2$ strings $x$ of length $n$ have complexity
$C(x)=n$ and at least $\Omega (2^n)$ strings $y$ of length $n$
have complexity $C(y)=n+1$. There are no strings $z$ of length $n$
that have complexity $C(z) > n+1$.
\end{remark}

\section{Average Time for Boolean Matrix Multiplication}~\label{sec.boolean}
We begin with a simple (almost trivial) illustration of average-case
analysis using the incompressibility method. Consider the problem
of multiplying two $n \times n$ boolean matrices $A = (a_{i,j})$ and
$B = (b_{i,j})$. Efficient algorithms for this problem have always been
a very popular topic in the theoretical computer science literature due to
the wide range of applications of boolean matrix multiplication. The best
worst-case time complexity obtained so far is $O(n^{2.376})$ due to 
Coppersmith and Winograd~\cite{CoWi87}. In 1973, O'Neil and O'Neil
devised a simple algorithm described below which runs in $O(n^3)$ time
in the worst case but achieves an average time complexity of
$O(n^2)$~\cite{oneil73}. 

\bigskip
{\noindent \bf Algorithm} QuickMultiply($A,B$)
\begin{enumerate}
\item Let $C = (c_{i,j})$ denote the result of multiplying $A$ and $B$.
\item For $i := 1$ to $n$ do
\item $~~~~$ Let $j_1 < \cdots < j_m$ be the indices such that $a_{i,j_k} = 1$,
      $1 \leq k \leq m$.
\item $~~~~$ For $j := 1$ to $n$ do
\item $~~~~~~~~$ Search the list $b_{j_1,j}, \ldots, b_{j_m,j}$ sequentially
      for a bit $1$.
\item $~~~~~~~~$ Set $c_{i,j} = 1$ if a bit $1$ is found, or $c_{i,j} = 0$
      otherwise.
\end{enumerate}

An analysis of the average-case time complexity of QuickMultiply 
is given in~\cite{oneil73} using simple probabilitistic arguments.
Here we give an analysis using the incompressibility method. 

\begin{theorem}\label{thm.bollean}
Suppose that the elements of $A$ and $B$ are drawn uniformly and
independently. Algorithm QuickMultiply runs in $O(n^2)$ time on the average. 
\end{theorem}
\begin{proof}
Let $n$ be a sufficiently large integer. Observe that the average time
of QuickMultiply is trivially bounded between $O(n^2)$ and $O(n^3)$.
By the Incompressibility Lemma, out of the $2^{2n^2}$ pairs of
$n \times n$ boolean matrices, at least $(n-1)2^{2n^2}/n$ of them are 
$\log n$-incompressible. Hence, it suffices to consider
$\log n$-incompressible boolean matrices.

Take a $\log n$-incompressible binary string $x$ of length $2n^2$, and form
two $n \times n$ boolean matrices $A$ and $B$ straightforwardly so that
the first half of $x$ corresponds to the row-major listing of the elements
of $A$ and the second half of $x$ corresponds to the row-major listing of
the elements of $B$. We show that QuickMultiply spends $O(n^2)$ time on
$A$ and $B$.

Consider an arbitrary $i$, where $1 \leq i \leq n$. It suffices to show that
the $n$ sequential searches done in Steps 4 -- 6 of QuickMultiply take a total
of $O(n)$ time. By the statistical results on various blocks in incompressible
strings given in Section 2.6 of~\cite{LiVi93}, we know that at least
$n/2 - O(\sqrt{n\log n})$ of these searches find a $1$ in the first step,
at least $n/4 - O(\sqrt{n\log n})$ searches find a $1$ in two steps,
at least $n/8 - O(\sqrt{n\log n})$ searches find a $1$ in three steps, and
so on. Moreover, we claim that none of these searches take more than
$4\log n$ steps. To see this, suppose that for some $j$, $1 \leq j \leq n$, 
$b_{j_1,j} = \cdots = b_{j_{4\log n},j} = 0$. Then we can encode $x$
by listing the following items in a self-delimiting manner:
\begin{enumerate}
\item A description of the above discussion.
\item The value of $i$.
\item The value of $j$.
\item All bits of $x$ except the bits $b_{j_1,j}, \ldots, b_{j_{4\log n},j}$.
\end{enumerate}
This encoding takes at most
\[ O(1) + 2\log n + 2n^2 - 4\log n + O(\log\log n) < 2n^2 - \log n \]
bits for sufficiently large $n$, which contradicts the assumption that
$x$ is $\log n$-incompressible.

Hence, the $n$ searches take at most a total of
\begin{eqnarray*}
&& (\sum_{k = 1}^{\log n} (n/2^k - O(\sqrt{n\log n})) \cdot k) + 
         (\log n) \cdot O(\sqrt{n\log n}) \cdot (4\log n) \\
& < & (\sum_{k = 1}^{\log n} kn/2^k + O(\log^2 n \sqrt{n\log n}) \\
& = & O(n) +  O(\log^2 n \sqrt{n\log n}) \\
& = & O(n) 
\end{eqnarray*}
steps. This completes the proof.   
\end{proof}

\section{Average Complexity of Finding the Majority}~\label{sec.majority}
Let $x = x_1 \cdots x_n$ be a binary string. The {\em majority bit}
(or simply, the {\em majority}) of $x$ is the bit ($0$ or $1$) that appears
more than $\lfloor n/2 \rfloor$ times in $x$. The majority problem is that,
given a binary string $x$, determine the majority of $x$. When 
$x$ has no majority, we must report so.

The time complexity for finding the majority has been well studied in the
literature (see, {\it e.g.}~\cite{AKL85,ARS93,ARS97,GK90,SW91}).
It is known that, in the worst case, $n - \nu(n)$ bit comparisons are
necessary and sufficient~\cite{ARS93,SW91}, where $\nu(n)$ is the 
number of occurrences of bit $1$ in the binary representation of number $n$.
Recently, Alonso, Reingold and Schott~\cite{ARS97} studied the average
complexity of finding the majority assuming the uniform probability
distribution model. Using quite sophisticated arguments based on
decision trees, they showed that on the average finding the majority
requires at most $2n/3 - \sqrt{8n/9\pi} + O(\log n)$ comparisons and
at least $2n/3 - \sqrt{8n/9\pi} + \Theta(1)$ comparisons.

In this section, we consider the average complexity of finding the majority
and prove a pair of upper and lower bounds tight up to the first major term,
using simple incompressibility arguments.

We start by proving an upper bound of $2n/3 + O(\sqrt{n\log n})$. 
The following standard tournament algorithm is needed.

\bigskip
{\noindent \bf Algorithm} Tournament($x = x_1 \cdots x_n$)
\begin{enumerate}
\item If $n = 1$ then return $x_1$ as the majority.
\item Elseif $n = 2$ then 
\item $~~~$ If $x_1 = x_2$ then return $x_1$ as the majority.
\item $~~~$ Else return ``no majority''.
\item Elseif $n = 3$ then 
\item $~~~$ If $x_1 = x_2$ then return $x_1$ as the majority.
\item $~~~$ Else return $x_3$ as the majority.
\item Let $y = \epsilon$.
\item For $i := 1$ to $\lfloor n/2 \rfloor$ do
\item $~~~$ If $x_{2i-1} = x_{2i}$ then append the bit $x_{2i}$ to $y$.
\item If $\lfloor n/2 \rfloor$ is even then append the bit $x_n$ to $y$.
\item Call Tournament($y$).
\end{enumerate}

\begin{theorem}\label{majority.upper}
On the average, algorithm Tournament requires at most
$2n/3 + O(\sqrt{n\log n})$ comparisons.
\end{theorem}
\begin{proof}
Let $n$ be a sufficiently large number.
Again, since algorithm Tournament makes at $n$ comparisons on any string of
length $n$, by the Incompressibility Lemma, it suffices to consider
running time of Tournament on $\log n$-incompressible strings.
Let $x = x_1 \cdots x_n$ be a $\log n$-incompressible binary string.
For any integer $m \leq n$, let $\sigma(m)$ denote the maximum number of
comparisons required by algorithm Tournament on any $\log n$-incompressible
string of length $m$.

We know from~\cite{LiVi93} that among the $\lfloor n/2 \rfloor$ bit pairs
$(x_1,x_2), \ldots, (x_{2\lfloor n/2 \rfloor -1},x_{2\lfloor n/2 \rfloor})$
that are compared in step 10 of Tournament, there are at least 
$n/4 - O(\sqrt{n\log n})$ pairs consisting of complementary bits. 
Clearly, the new string $y$ obtained at the end of step 11 should satisfy
\[ C(y) \geq l(y) - \log n - O(1) \]
Hence, we have the following recurrence relation for $\sigma(m)$:
\[ \sigma(m) \leq \lfloor m/2 \rfloor + \sigma(m/4 + O(\sqrt{m\log n})) \]
By straightforward expansion, we obtain that
\begin{eqnarray*}
\sigma(n) & \leq &  \lfloor n/2 \rfloor + \sigma(n/4 + O(\sqrt{n\log n})) \\
& \leq &  n/2 + \sigma(n/4 + O(\sqrt{n\log n})) \\
& \leq &  n/2 + (n/8 + O(\sqrt{n\log n})/2) + 
      \sigma(n/16 + O(\sqrt{n\log n})/4 + O(\sqrt{(n\log n)/4})) \\
& = &  n/2 + (n/8 + O(\sqrt{n\log n})/2) + 
      \sigma(n/16 + (3/4) \cdot O(\sqrt{n\log n})) \\
& \leq &  \cdots  \\
& \leq &  2n/3 + O(\sqrt{n\log n})
\end{eqnarray*}
\end{proof}

Now we prove a lower bound which differs from the above upper bound
only by $O(\sqrt{n\log n})$.

\begin{theorem}\label{majority.lower}
Every algorithm requires at least $2n/3 - O(\sqrt{n\log n})$ comparisons
to find the majority, on the average.
\end{theorem}
\begin{proof}
Consider an arbitrary majority finding algorithm $A$.
Again, let $n$ be a sufficiently large number and $x = x_1 \cdots x_n$ a 
$\log n$-incompressible binary string. Without loss of generality,
we assume that $A$ never makes redundant comparisons, {\it i.e.}
if the relationship between bits $x_i$ and $x_j$ can be inferred from
the previous comparisons, then $A$ will not compare $x_i$ with
$x_j$ again. It will be useful to think of the comparisons of $A$
as partitioning the bits $x_1, \ldots, x_n$ of $x$ into ``clusters''
where each cluster contains all the bits whose relationships to each
other have been established. Let $C_1, \ldots, C_p$ be the clusters
formed when $A$ terminates. For each cluster $C_i$, let $w(C_i)$,
called the weight of $C_i$,
denote the absolute value of the difference between the number of $0$'s
and the number of $1$'s in $C_i$. Clearly, in order for algorithm $A$
to be correct, there must be a unique cluster $C_i$ such that
\[ w(C_i) > \sum_{j \neq i} w(C_j). \]
(Otherwise how can the algorithm declare the majority)

We first claim that for each cluster $C_i$, its weight 
$w(C_i) \leq O(\sqrt{n\log n})$.
Suppose that $C_i$ contains $m$ bits $x_{i_1}, \ldots, x_{i_m}$,
listed in the order that they were first compared. 
Clearly, given the algorithm $A$, the bits in the clusters
$C_1, \ldots, C_p$, listed in the order that they were first compared,
encode the string $x$. Since the total number of the bits is $n$ and
$C(x) \geq n - \log n$, we have 
$C(x_{i_1} \cdots x_{i_m}) \geq m - \log n - O(1)$.
Hence, from the results in~\cite{LiVi93} we know that the numbers of $0$'s
and $1$'s in these bits differ by at most 
$O(\sqrt{m\log n}) \leq O(\sqrt{n\log n})$.

Suppose that $A$ makes a total of $k$ comparisons.
In order to relate $k$ to the Kolmogorov complexity of the string $x$,
we re-encode $x$ by listing the following information in the
self-delimiting form:

\begin{enumerate}
\item The above discussion and the algorithm $A$.
\item For each of the $k$ comparisons made by $A$, a bit indicating the outcome
      of the comparison. This gives rise to a string $y$ of length $k$.
\item For each cluster $C_i$, a bit indicating the value
of the lowest indexed bit in $C_i$.
\end{enumerate}

Since $A$ does not make redundant comparisons, the length of the above
description is at most $n + \log n + O(1)$.  Hence,
\[ k + p = l(y) + p = n \pm \log n \pm O(1) \] 
Thus, $C(y) \geq k - 2\log n$. Again, by the results in~\cite{LiVi93}, we
claim that at most $k/2 + O(\sqrt{k\log n})$ of the $k$ comparisons
identify pairs of complementary bits.

Observe that for any cluster $C_i$, in order for the weight $w(C_i)$ to equal 
zero, at least one of the comparisons that form $C_i$ must involve
complementary bits. Hence, from the above discussion, the number of clusters
with zero weight is at most $k/2 + O(\sqrt{k\log n})$. 
Since the maximum weight of a cluster is $O(\sqrt{n\log n})$, we have
\[ p - k/2 - O(\sqrt{k\log n}) \leq O(\sqrt{n\log n}) \]
Since $p \geq n - k - O(1)$, we obtain
\[ n - k - O(1) - k/2 - O(\sqrt{k\log n}) \leq O(\sqrt{n\log n}) \]
That is,
\[ k \geq 2n/3 - O(\sqrt{k\log n}) - O(\sqrt{n\log n}) =
   2n/3 - O(\sqrt{n\log n}). \]
\end{proof}

\section{Communication Complexity}
Consider the following communication complexity problem (for  
definitions see the book by Kushilevitz and 
Nisan~\cite{KushilevitzNisan97}). Initially, Alice has a string $x = 
x_1,\ldots,x_n$ and Bob has a string $y = y_1,\ldots,y_n$ with
$x,y \in \{0,1\}^n$. Alice and Bob
use an agreed-upon protocol 
to compute the inner product of $x$ and $y$ modulo 2
\[f(x,y) = \sum_{i=1}^n x_i . y_i \bmod 2\]
with Alice ending up with the result.
We are interested in the minimal possible number 
of bits used in communication between Alice and Bob in such a protocol. 
Here we prove a lower bound of $n-1$, which is almost 
tight since the trivial protocol 
where Bob sends all his $n$ bits to Alice achieves this bound. 
In ~\cite{KushilevitzNisan97} it is shown that the lower bound is 
in fact $n$. We also show an $n-O(1)$ lower bound for the average-case
complexity and a $n-1 + \log (1- \epsilon )$ lower bound
for randomized algorithms (private coins) that output the correct
answer with probability at least $1 - \epsilon$.

\begin{theorem}
Assume the discussion above. Every protocol computing
the inner product function requires at least $n-1$ bits of communication.
\end{theorem}
\begin{proof}
Fix a communication protocol $P$ that computes the inner product.
Let $A$ be an algorithm that we describe later.
Let $z$ be a string of length $2n$ such that $C(z| A, P,n) \geq 2n-1$.
Let $z = x_1\ldots x_ny_1\ldots y_n$ Let Alice's input be  $x =
x_1\ldots x_n$ and Bob's input be $y_1\ldots y_n$. Assume  without
loss of generality that $f(x,y) =
0$ (the innerproduct of $x$ and $y$ is $0$ modulo $2$).
\footnote{By symmetry there are precisely $2^{2n-1}$ strings $z=xy$
$l(x)=l(y)=n$ with inner product $x,y$ equal 0. There are only
$2^{2n-1}-1$ programs of length less than $2n-1$. Hence there must be
a $z$ as required.}
Run the communication protocol $P$ 
between Alice and Bob ending in a state where Alice outputs that $f(x,y)$ is
$0$. 
Let $C$ be the sequence
of bits sent back and forth.
Note that $P$ can be viewed as a tree with $C$ a path in this
tree~\cite{KushilevitzNisan97}. Hence $C$ is self-delimiting.
Consider the set $S$ defined by
\[
S := \{ a : \exists b \mbox{ such that } P(a,b) = 0 \mbox{ and
induces conversation C }, a,b \in \{0,1\}^n \}.
\]
Given $n, P$ and $C$, we can compute $S$.
Let the cardinality of $S$ be $l$. The strings in $S$
form a matrix $M$ over GF(2) with the $i$th row of $M$ 
corresponding to the $i$th
string in $S$ (say in lexicographic ordering). Since for every $a \in
S$ it holds that $f(a,y) = 0$ it follows that $y$ is an element of the Null
space of $M$ ($y \in \mbox{Null}(M)$). Application of the Null space Theorem
from linear algebra yields:
\begin{equation}\label{nullspace}
\mbox{rank}(M) + \mbox{dim(Null(M))} = n.
\end{equation}
Since the cardinality of $S$ is $l$ and we are working over GF(2) it follows
that the rank of $M$ is at least $\log(l)$ and by (\ref{nullspace}) it follows
that dim(Null($M$)) $\leq n - \log(l)$.  The following is an
effective description of $z$ given $n$ and the reconstructive algorithm $A$
explained below:

\begin{enumerate}
\item $C$;
\item the index of $x \in S$ using $\log(l)$ bits; and
\item the index of $y \in$ Null($M$) with $n - \log(l)$ bits.
\end{enumerate}

The three items above can be concatenated without delimiters. Namely,
$C$ itself is self-delimiting, while from $C$ one can generate $S$ and hence 
compute $l$.  From the latter item one can compute the binary length
 of the index for 
$x \in S$, and the remaining suffix
of the binary description is the index for $y \in$ Null($M$). 
 From the given description and $P,n$ the algorithm $A$ reconstructs $x$ and $y$
and outputs $z=xy$. Consequently, $C(z|A,P,n) \leq l(C) + \log l + (n- \log l)$.
Since we have assumed $C(z|A,P,n) \geq 2n-1$
it follows that $l(C) \geq n-1$.    
\end{proof}

We can improve this lower bound to $n$ as follows:

\begin{theorem}
Assume the discussion above. There exists a constant $c$
such that for all $m$ there is an $n$ ($2m-c \leq 2n \leq 2m$) 
such that every protocol computing
the inner product function of two $n$-bit strings
requires at least $n$ bits of communication.
\end{theorem}
\begin{proof}
Using a similar argument as in Theorem~\ref{theo.nurand2} there
is a constant $c$ such that for every $m$
we can choose $z$ of
length $2n$ ($2m-c \leq 2n \leq 2m$)
with associated inner product 0 and $C(z|A,P,n) \geq 2n$.
The remainder of the proof is the same as above.
\end{proof}

Approximately the same lower bound holds for the average-case
communication complexity
of computing the inner product of two $n$-bit strings:
\begin{theorem}
The average communication complexity
of computing the inner product of two $n$-bit strings is 
at least $n-O(1)$ bits.
\end{theorem}
\begin{proof}
There are exactly $2^{2n-1}$ strings $z$ of length $2n$ such that
$z=xy$, $l(x)=l(y)=n$ and inner product of $x$ and $y$ modulo 2 equals
1. For such $z$ define $\bar{z} = \bar{x} \bar{y}$ as $z$ but
with the first bit of $x$ and the first bit of $y$ changed 
so that the inner product of the resulting strings $\bar{x}$
and $\bar{y}$
equals 0.
$C( \bar{w} |n)= C(w|n) + O(1)$ for every $w \in \{z,x,y\}$.
Let $\delta (n)$ be a function and choose in the proof above
$C(z|A,P,n) \geq 2n - \delta (n)$. By simple counting
there are at least 
\begin{equation}\label{eq.sc}
2^{2n}(1 - 1/2^{\delta (n)}) 
\end{equation}
such $z$'s.
If the above inner product associated with $z$ equals 1 then 
the inner product associated with $\bar{z}$ equals 0
and $C(\bar{z}|A,P,n) \geq 2n - \delta (n)-O(1)$

Hence we can apply the proof of the previous theorem
for all $z$ with randomness deficiency at most $\delta (n)$ as follows:
\begin{itemize}
\item If $z$ has an associated inner product 0 then the proof as above
yielding $l(C_z) \geq n - \delta (n)$ where $C_z$ is the communication
sequence associated with the computation with input $z$.
\item If $z$ has associated inner product 1 then the proof as above
to $\bar{z}$ yielding $l(C_{\bar{z}}) \geq n - \delta (n) - O(1)$. 
\end{itemize}
There are at least $2^{2n}(1-1/2^{\delta (n)})$ strings $z$
of length $2n$ with $C(z|A,P,n) \geq 2n - \delta (n)$.
Altogether we obtain that the average communication complexity is
\begin{eqnarray*}
\sum_{z \in \{0,1\}^{2n}} {\rm Pr}(z) l(C_z) & = &
2^{-2n}  \sum_{z \in \{0,1\}^{2n}} l(C_z)\\
& \geq & 
2^{-2n} \sum_{\delta (n)=1}^{n} 
\sum_{z \in \{0,1\}^n \& C(z|A,P,n) = 2n - \delta (n)} l(C_z)\\
& \geq &
2^{-2n} \sum_{\delta (n)=1}^n 2^{2n} \frac{1}{2^{\delta (n)}}(n-\delta (n)-O(1))\\
& = &
\sum_{\delta (n) = 1}^n \frac{n}{2^{\delta (n)}} - \sum_{\delta
(n)=1}^n \frac{\delta (n)+O(1)}{2^{\delta (n)}}\\
& \geq & n - O(1).
\end{eqnarray*}


This proves the theorem.
\end{proof}

A similar proof establishes a lower bound of about
$n$ bits for the communication complexity
of equality of the strings held by both parties.

\begin{theorem}
The communication complexity of a randomized protocol using
private coins 
computing the inner product of two $n$-bit strings 
that outputs the correct answer with probability at least $1- \epsilon$
is at least $n-1 + \log (1- \epsilon )$ bits.
\end{theorem}
\begin{proof}
The  proof is similar to that of the determinsitic lower bound. 
Assume that Alice and Bob compute probabilistically, 
that is, they can each flip a private fair coin (whoes output does not
depend on the input) and decide their next step
depending on the result of the coin flip, and the
error rate of their computation is $\epsilon$. 
Because for each pair of inputs, there is $1-\epsilon$ chance to
output the correct result, there must exist a coin flipping sequence
such that using it 
Bob and Alice output the correct result for $1-\epsilon$ portion of the inputs.
Fix such a sequence $R$. Out of such $2^{2n} (1-\epsilon )$ strings of
length $2n$, choose $z$ such that
$$
C(z|A,P,R,n)>= 2n -1 + \log (1-\epsilon).
$$
Such $z$ exists according to Lemma~\ref{C2}. Then we procede as
before in the edeterministic case to show that Alice and Bob must communicate 
$$
n -1 + \log (1-\epsilon )
$$
bits to compute $f(x,y)$, where $z=xy$, $|x|=|y|=n$.
\end{proof}

\section{Acknowledgements}
We thank Ian Munro for discussions on related subjects, 
and Bill Smyth for
introducing us to the paper~\cite{ARS97}.

\end{document}